\documentstyle[amssymb,12pt]{article}

\input{tcilatex}

\begin{document}

\begin{titlepage}
\renewcommand{\baselinestretch}{1}
\renewcommand{\thepage}{}
\title{\bf A Model for the Schottky  Anomaly\\
in Metallic $Nd_{2-y}Ce_{y}CuO_{4}$} 
\author{R. E. Lagos$^{(a)}$, A. C. M. Stein-Barana$^{(a)}$ and
G. G. Cabrera$^{(b)}$\\
$(a)$ Departamento  de  F\'{\i}sica- IGCE  \\
Universidade Estadual Paulista (UNESP)\\               
C.P. 178,  Rio Claro 13500-970, SP, Brazil\\
$(b)$ Instituto de F\'{\i}sica $Gleb$ $Wataghin$\\
Universidade Estadual de Campinas (UNICAMP)\\
C.P. 6165, Campinas 13083-970, SP, Brazil }
    
\date{}
\maketitle

\begin{abstract} 
We present a simple model for the doped compound 
$Nd_{2-y}Ce_{y}CuO_{4}$, in order to explain some recent
experimental results on the latter.  Within a Hartree-Fock context,
we start from an impurity Anderson-like model and consider the
magnetic splitting of the $Nd$-$4f$ ground state Kramers doublet
due to exchange interactions with the ordered $Cu$ moments. Our
results are in very good agreement with the experimental data, 
yielding a Schottky anomaly peak for the specific heat that
reduces its amplitude, broadens and shifts to lower temperatures, upon $Ce$
doping. For overdoped compounds at low temperatures, the specific heat
behaves linearly and the magnetic susceptibility is constant. A smooth transition 
from this Fermi liquid like behavior ocurrs as temperature is increased and at high 
temperatures the susceptibility  exhibits a Curie-like behavior. Finally, we discuss
some improvements our model is amenable  to incorporate.

\ 

\

\begin{flushleft}
{\em PACS codes/keywords}: 65.40.Hq, 75.20.Hr, 74.72.Jt\\
Schottky Anomalies, Rare Earths, Anderson Model, Cuprate Oxides\\
{\em Corresponding author}: R. E. Lagos\\
Departamento  de  F\'{\i}sica- IGCE, Universidade Estadual Paulista (UNESP)\\
C.P. 178, Rio Claro 13500-970, SP, Brazil\\
fax: (55)-534-8250 \hspace{0.2in} email:monaco@laplace.igce.unesp.br
\end{flushleft}
\end{abstract}

\end{titlepage}

\newpage We develop a model for the compound $Nd_{2-y}Ce_{y}CuO_{4}$,
following both the conjectures and the experimental results as given by \cite
{czj}. There, a Schottky-like peak in the specific heat curve $C(T)$ was
observed. As $Ce$ doping is increased, this peak reduces its amplitude,
broadens and shifts to lower temperatures. In the overdoped case, at low
temperatures, $C(T)$ behaves linearly and the magnetic susceptibility
remains constant. At high temperatures, the former behaves exponentially and
the latter in a Curie-like fashion. A large effective mass was inferred and
interpreted as a heavy fermion feature. Czjzek el al. \cite{czj}
interpretation is as follows: the Schottky-like peaks result from the
magnetic splitting of the $Nd$-$4f$ ground state Kramers doublet due to
exchange interactions with the ordered $Cu$ moments. Taking at face value
their interpretation and data, we develop a simple model for this compound's
behavior.

We start with an impurity Anderson-like model \cite{ander}, within a
Hartree-Fock ($HF$) context, and follow a somehow similar modeling as in 
\cite{sim}. We denote by $\varepsilon _k$ and $n_{k\sigma }=c_{k\sigma
}^{\dagger }c_{k\sigma }$ the conduction band dispersion and occupation
number operator, respectively. For the undoped case the Fermi level (taken
as zero throughout this calculation) lies in the gap below the bottom of the
conduction band (the undoped compound exhibits semiconducting behavior and
undergoes a metal-insulator transition at $y=0.13$ \cite{czj}). As the
system is doped with electrons \cite{brian} the band `moves' towards the
Fermi level (we assume the metal insulator transition of the band crossing
type \cite{ehm}). In the present modeling, we only consider the doublet $f$%
-electron state structure, labeled by the spin variable $\sigma $ (orbital
degeneracy and crystal field effects are neglected at the present time). Let 
$n_{fi\sigma }=f_{i\sigma }^{\dagger }f_{i\sigma }$ be the localized $f$%
-electron level occupation number, at energy $E_o$. The $f$ levels are
assumed singly occupied (equivalent to consider the $f$-intra atomic Coulomb
term $U_{ff}\rightarrow \infty $). Furthermore we denote by $J$ the exchange
interaction between the $f$ and the conduction electrons. The $Cu$ moment
coupling via exchange to the $f$-electrons is modeled in a {\it %
non-selfconsistent} fashion, and the $Cu$ moment mean value {\it times} the
exchange constant is denoted by $D$. Let $h$ be the external magnetic field,
and the conduction ($c$) and $f$-electron magnetic moment mean values be
denoted by $m_{c,f}$, respectively (absorbing all magnetic constants).
Finally, the conduction-$f$ electron hybridization is modeled via the
Anderson impurity model \cite{ander}, so the $HF$ Hamiltonian for this model
is cast as

\ 

\begin{equation}
H=\sum_{k\sigma }E_{k\sigma }n_{k\sigma }+\sum_{i\sigma }E_{f\sigma
}n_{fi\sigma }+\frac{V}{\sqrt{N}}\stackunder{ki\sigma }{\sum }\left(
c_{k\sigma }^{\dagger }f_{i\sigma }+f_{i\sigma }^{\dagger }c_{k\sigma
}\right)   \label{ham}
\end{equation}

\ 

\noindent except for a constant term and where $V$ is the hybridization
constant, $N$ the total number of sites, and

\ 

\[
E_{k\sigma }=\varepsilon _{k}-\sigma (Jm_{f}+h)\hspace{0.5in}E_{f\sigma
}=E_{o}-\sigma (D+Jm_{c}+h) 
\]

\ 

Notice the similarity of this Hamiltonian with the one presented by \cite
{sim} in a related but somehow different context. Following the standard $HF$
approach, as given by \cite{ander}, we obtain the selfconsistent equation
for the $f$-moment

\ 

\begin{equation}
m_{f}=-\frac{Im}{\pi }\sum_{\sigma }\sigma \int \frac{d\omega f_{T}(\omega )%
}{\omega -E_{f\sigma }+i\Delta (\omega )}  \label{mom}
\end{equation}

\ 

\noindent with $f_{T}(\omega )$ the Fermi-Dirac distribution and the $f$%
-level broadening factor $\Delta $ is a function of the energy $\omega $,
the hybridization constant $V$ and the conduction electron density of states
as given by $\Delta (\omega )=\pi V^{2}\rho _{c}(\omega )$ \cite{ander}.
Notice the $f$-moment is normalized to the effective moment $\mu _{eff}$ 
\cite{newns}. For a wide band metal, $\Delta $ is well approximated by its
value at the Fermi level. Here, this is not the case since a metal insulator
transition occurs as the compound is doped. We approximate $\Delta $ by its
energy average (around the Fermi level), as $\Delta (\omega )\approx \Delta
(y)=\pi V^{2}\bar{\rho}_{c}(y)$. This `average' is a monotonic function of $%
Ce$ doping. The conduction band moment can be likewise approximated as $%
m_{c}\approx 2(Jm_{f}+h)\bar{\rho}_{c}(y)$. Once the $f$ -moment is computed
(selfconsistently), we are in a position to compute the $f$ contribution to
the energy as

\ 

\begin{equation}
{\cal E}=-\frac{Im}{\pi }\sum_{\sigma }\int \frac{d\omega \omega
f_{T}(\omega )}{\omega -E_{f\sigma }(y)+i\Delta (y)}  \label{energy}
\end{equation}

\ 

The specific heat and magnetic susceptibility are readily computed, at zero
magnetic field, respectively as

\ 

\begin{equation}
{\cal C}=\frac{\partial {\cal E}}{\partial T}\hspace{0.5in}\chi =\left( 
\frac{\partial m_{f}}{\partial h}\right) _{h=0}  \label{def}
\end{equation}

\ 

Our results are summarized in the figures. In Figure $1$ and $2$ we plot the
specific heat (in arbitrary units) and the specific heat over temperature,
respectively; in Figure $3$ the $f$ magnetic susceptibility and in Figure $4 
$ the $f$ magnetic moment; all versus temperature in log scale. We attempted
a quantitative fitting to the specific heat experimental curves \cite{czj}.
In each Figure we plot four cases, corresponding to the parameters'
configurations (in Kelvin units) listed in the Table, with $y$ the $Ce$
doping relative concentration. The effective mass is estimated via the
expression

\ 

\[
\rho _f(\varepsilon _F\cong 0,T=0)=\frac{\Delta (y)}{\Delta ^2(y)+\left(
D(y)+J(y)m(y)\right) ^2} 
\]

\[
\frac{m^{*}}m\cong \frac{2\hbar ^2}{ma^2}\rho _f(\varepsilon _F\cong 0,T=0) 
\]

\ 

\noindent where $a$ is the n.n. $Nd$ lattice constant, and taken from \cite
{lattice}. In the last expression we fit the Lorentzian $f$-broadening with
an effective tight binding Hamiltonian density of states.\cite{tbh}.
Furthermore, we consider $E_o-\varepsilon _F\approx 0$ (a non selfconsistent
approximation, used in our model to approximately `force' a singly occupied $%
f$ -level). The parameters are computed within a virtual crystal
approximation scheme \cite{ehm, corr, corr1}. This linear interpolation
scheme mimics the effect $Ce$ doping, thus $\Delta $ must increase with
doping, as electron injection enhances the conduction density of states at
the Fermi level. Correspondingly, electron doping must thin down the
`magnetic' parameters $D$ and $J$, respectively associated to the incomplete
filling of the $Cu$ and $Nd$ orbitals.

From Figures $1$ and $2$ we notice the overall linear behavior of the
specific heat at low temperatures, and an exponential behavior at high
temperatures, in agreement with the experimental data \cite{czj}. Also, as
doping ($\Delta $) is increased, the Schottky-like peak moves to lower
temperatures, reduces its amplitude and broadens, again in agreement with
the experimental data. Notice that within our crude modeling for $\Delta $,
a non linear behavior at low temperatures is evident only for near null
values of $\Delta $, nevertheless our modeling stands as a good description
for the metallic phase of this rare earth compound (this last remark is also
valid for the computed susceptibility). $\Delta (\omega ,y)$ is a function
of both energy and doping, in fact proportional to the conduction band
density of states $\rho _{c}(\omega ,y)$. For the undoped case (insulator) $%
\rho _{F}=\rho _{c}(\omega =\varepsilon _{F},y)=0$. As doping is increased, $%
\rho _{F}$ becomes finite at the insulator-metal transition. Consider our
expression for the specific heat (equations (\ref{energy}) and (\ref{def}))
with the energy independent approximation $\Delta (y)$. For the case $T\ll
\Delta $, the resulting Digamma functions can be expanded (Sommerfeld
expansion) rendering the well known linear expression ${\cal C}=\gamma T$.
On the other hand, for the case $\Delta =0$ and with $\varepsilon =\left|
E_{f\uparrow }-E_{f\downarrow }\right| $, it can be readily shown that

\[
{\cal C}\approx \left( \frac{\varepsilon }{T}\right) ^{2}\frac{\exp \left( 
\frac{\varepsilon }{T}\right) }{\left( 1+\exp \left( \frac{\varepsilon }{T}%
\right) \right) ^{2}} 
\]

\noindent the classical Schottky expression (see for example \cite{kubo}),
with an exponential behavior at low temperatures and a sharp maximum at $T$ $%
\approx \varepsilon .$ So as $\Delta $ varies away from zero we expect a
transition from an exponential to a linear behavior. Thus, our single
particle model accounts for the above mentioned transition as can be
inferred from the experimental curves \cite{czj} . The approximation $\Delta
(\omega ,y)\approx \rho _{F}$ is a {\it good approximation} irrespective of
the value of $\rho _{F}$ {\it provided} $\rho _{c}(\omega ,y)$ is a {\it %
slowly varying function}. For low doping this is not the case (an
insulator-metal transition takes place), therefore if an energy independent
approximation should be made for $\Delta $, equating the latter to $\rho
_{F} $ turns out to be a naive guess. Nevertheless qualitative behavior is
obtained; namely the exponential to linear behavior in the low temperature
region for the specific heat, and the maximum peak of the latter reduces its
amplitude, broadens and shifts to lower temperatures as doping ($\Delta $)
is increased. We chose the effective (energy independent) $\Delta ^{\prime
}s $ to best fit the experimental data on the specific heat, specially with
regard to the peaks' height and width. The low temperature regime for the
underdoped cases exhibits a minute non linear behavior (see Figs.$1$ and $2$%
). Detailed analysis of our results shows a very weak exponential behavior
`creeping in' as doping is reduced. We believe this quantitative shortcoming
at both low temperature and doping is unavoidable within the effective $%
\Delta $ approximation. This situation can be remedied by considering in
earnest a `rapidly' varying $\Delta (\omega ,y)$.

Figure $3$ displays the moment suppression as doping and/or temperature are
increased. Figure $4$ displays the Pauli-like behavior of the system at low
temperatures, and a Curie-like behavior at high temperatures. This feature
agrees with the experimental data, for the metallic case $y=0.20$ ($%
m^{*}\approx 1000m$).

From Figures $2$ and $4$, and only for case $d$ we notice a smooth
transition from a Fermi liquid behavior (with a large effective mass $%
\approx 1000m$) to a (two level) free-ion behavior at high temperatures. All
other cases ($m^{*}\lessapprox 600m$), and for which no data was presented
in \cite{czj}, exhibit a maximum at intermediate temperatures, suggesting an
intermediate valence profile \cite{newns}.

The specific heat peak occurs at the same temperature where the moment is
suppressed. Its amplitude is related in a monotonic fashion to the maximum
moment value. This feature cannot be understood by just tuning the $f$-level
splitting. Thus, the doping dependent hybridization mechanism seems to be an
essential ingredient to explain the observed Schottky-like phenomena.

Two distinct regimes seem to be present. The first one, hereafter denoted
intermediate valence regime (IVR, composed of the a, b and c cases, with $%
m^{*}\lessapprox 600$), and the second, associated to case $d$ ($%
m^{*}\approx 1000m$) with no susceptibility peak, as seen from Figure $4$.
Following \cite{czj} we identify it as belonging to a heavy fermion regime
(HFR). Both regimes can be discriminated by their effective mass ratio and
by their respective susceptibility profile (Figure $4$). There, the IVR
shows a maximum in the region separating the Pauli behavior (low $T$) from
the Curie behavior (high $T$). In our model calculation such a maximum is
smeared out in the HFR and the $f$-moment is further reduced as we cross
over from the IVR to the HFR. (see Figure $3$). Consistent with this
picture, spin fluctuations are large at the Schottky peak temperature for
the IVR (see Figure $4$). In contrast, the HFR exhibits, as temperature is
lowered , monotonically increasing spin fluctuations, needed for the $f$%
-moment reduction.

We conclude that in our model the $Cu$ moment exchange with the $f$%
-electrons is responsible for the appearance of the Schottky-like peaks. The
peaks' structure are driven by doping via a simple Anderson mechanism. The
latter also shapes the linear behavior of the specific heat and the
Pauli-like susceptibility, at low temperatures. As doping is increased the
system seems to cross from an intermediate valence to a heavy fermion
regime. All these features are in good agreement with the experimental data
presented in ref. \cite{czj}.

Our model stands as a simple and relevant analysis of this Schottky like
effect, within a Hartree-Fock scheme. We believe that in order to explore
further results of this model, such as the temperature scale separating the
Fermi liquid-like behavior from the high temperature free-ion behavior, or
to attempt further quantitative fitting with experimental results, we must
first incorporate correlation effects in earnest, and lift some of the
approximations used in our model. We briefly mention some improvements our
model is amenable to incorporate. We should go beyond $HF$, as in \cite{ehm},%
\cite{corr},\cite{corr1},\cite{exact}-\cite{sabine} so the $%
U_{ff}\rightarrow \infty $ limit is adequately treated, as for example with
an effective medium approach \cite{ehm},\cite{corr}-\cite{corr1}. Within
this scheme, both non-selfconsistent restrictions used in our model namely: $%
E_{o}-\varepsilon _{F}\approx 0$ and to consider the $Cu$ moment as a
`classical' parameter ($D$), can be readily lifted and treated in earnest.
Furthermore, the approximation for $\Delta (\omega ,y)$ as solely a doping
dependent function can be relaxed. See for example the phenomenological
scheme in \cite{jr} and the renormalization group approach in \cite{renor}.
These above mentioned improvements will be considered in a forthcoming paper.%
\vspace{0.5in}

\noindent {\it Acknowledgments}: This work was partially supported by CNPq,
Brazil, and FAPESP, SP, Brazil.

\newpage\ 

\begin{center}
{\bf Table}

\vspace{0.5in}

\begin{tabular}{|c|c|c|c|c|c|}
\hline
& $y$ & $\Delta $ $[^{o}K]$ & $D$ $[^{o}K]$ & $J$ $[^{o}K]$ & $\frac{m*}{m}%
\times 10^{3}$ \\ \hline
a & $0.00$ & $0.90$ & $0.50$ & $3.50$ & $0.23$ \\ \hline
b & $0.10$ & $1.20$ & $0.38$ & $2.45$ & $0.34$ \\ \hline
c & $0.15$ & $1.35$ & $0.31$ & $1.93$ & $0.57$ \\ \hline
d & $0.20$ & $1.50$ & $0.25$ & $1.40$ & $1.14$ \\ \hline
\end{tabular}

\vspace{0.5in}\vspace{0.5in}

{\bf Figure Captions}\vspace{0.5in}
\end{center}

\begin{description}
\item  Figure {\bf 1}: Specific heat (arbitrary units) versus temperature.
Four parameters configurations in Kelvin units, labeled a,b,c and d,
respectively (see Table 1).

\item  Figure {\bf 2}: Specific heat (arbitrary units) over temperature,
versus temperature.

Labeling follows Fig. 1.

\item  Figure {\bf 3}: $f$-magnetic moments versus temperature. Labeling
follows Fig. 1.

\item  Figure {\bf 4}: $f$-magnetic susceptibility versus temperature.
Labeling follows Fig. 1
\end{description}

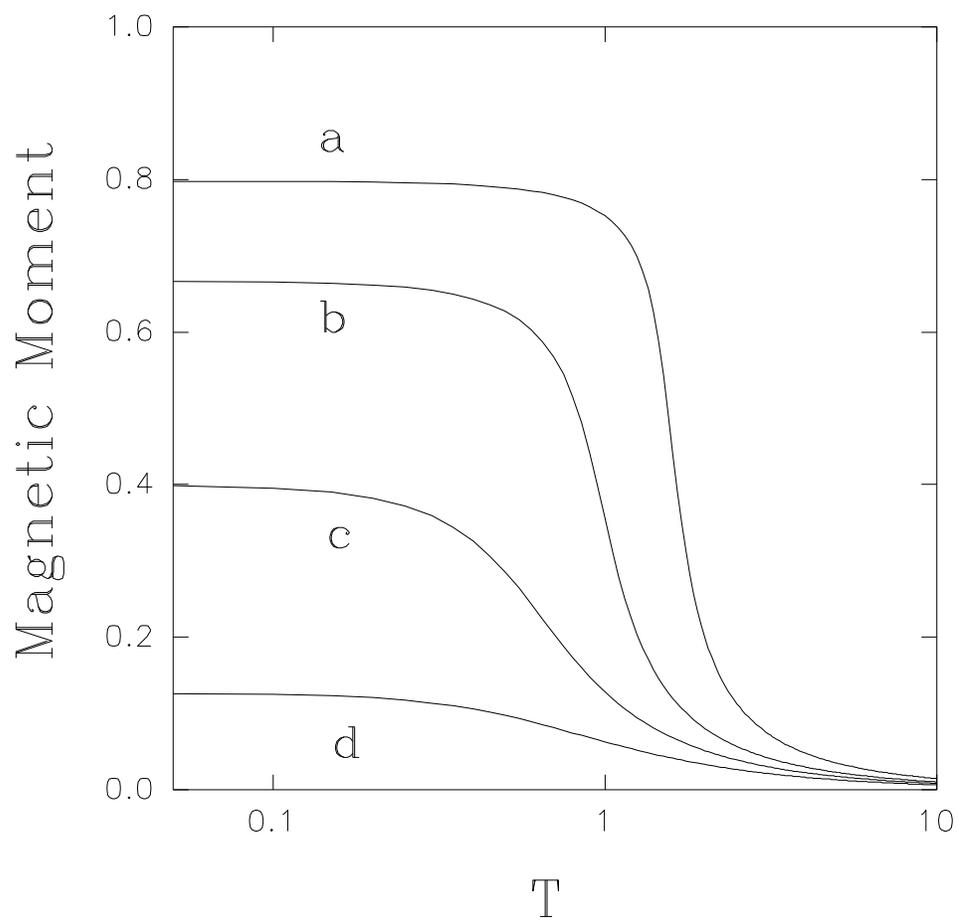
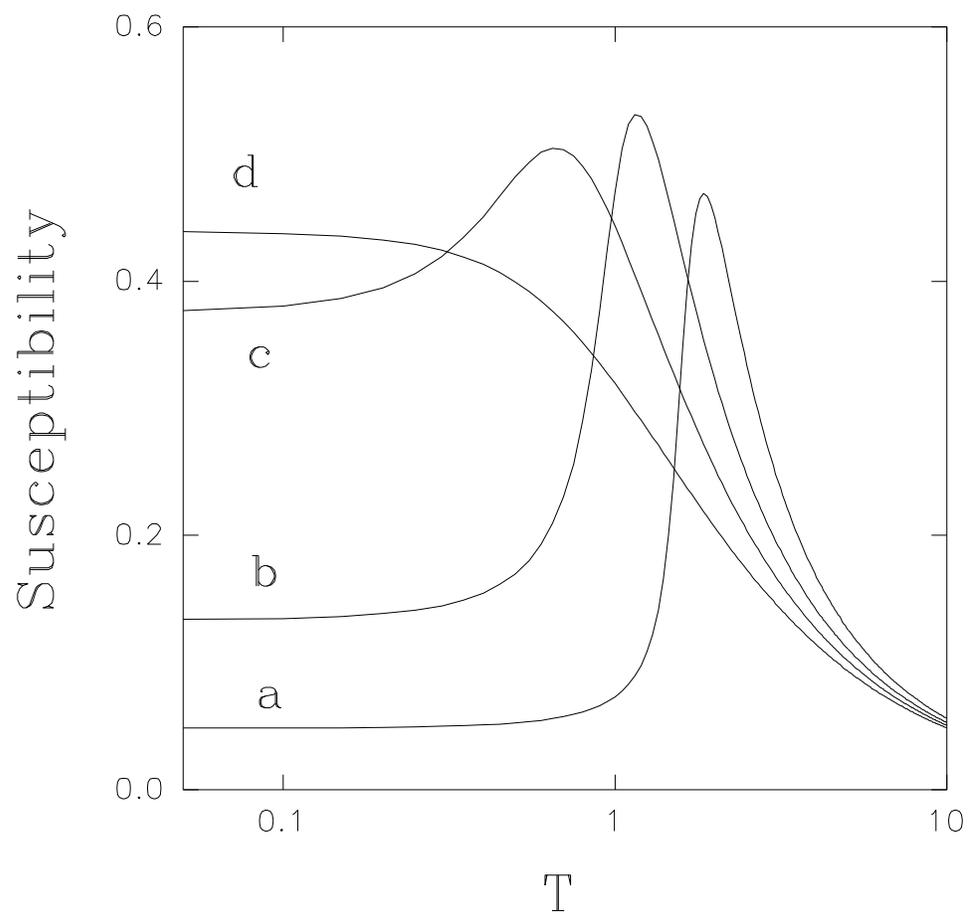
\end{document}